 \documentclass[preprint]{aastex63}

\usepackage{natbib}
\usepackage{diagbox}

\submitjournal{ApJ}

\shorttitle{The effects of Kappa electrons on \ion{O}{2} lines}
\shortauthors{Lin \& Zhang}
\begin{document}

\title{Can the Kappa-distributed electron energies account for the
intensity ratios of \ion{O}{2} lines in photoionized gaseous nebulae?}

\correspondingauthor{Yong Zhang}
\email{zhangyong5@mail.sysu.edu.cn}

\author{Bao-Zhi Lin}
\affiliation{School of Physics and Astronomy, Sun Yat-sen University, 
Zhuhai, 519082, China}

\author[0000-0002-1086-7922]{Yong Zhang}
\affiliation{School of Physics and Astronomy, Sun Yat-sen University, 
Zhuhai, 519082, China}

%% Note that the \and command from previous versions of AASTeX is now
%% depreciated in this version as it is no longer necessary. AASTeX 
%% automatically takes care of all commas and "and"s between authors names.

%% Mark off the abstract in the ``abstract'' environment. 
\begin{abstract}

A vexing puzzle in the study of planetary nebulae and \ion{H}{2} regions is that the plasma diagnostic results based on collisionally excited lines systematically differ from those based on recombination lines. A fairly speculative interpretation is
the presence of nonthermal electrons with the so-called $\kappa$ energy distributions, yet there is little observational evidence to verify or disprove this hypothesis. In this paper, we examine the influence of $\kappa$-distributed electrons  on the emissivities of \ion{O}{2} recombination lines 
using an approximate method, where the rate coefficients for a $\kappa$ distribution are computed
by summing Maxwellian-Boltzmann rate coefficients with appropriate weights.
The results show that if invoking $\kappa$-distributed electrons, the temperatures derived from the [\ion{O}{3}] $(\lambda4959+\lambda5007)/\lambda4363$ ratios could coincide with those estimated from the \ion{O}{2}  $\lambda4649/\lambda4089$ ratios. 
However, the estimated temperatures and $\kappa$ values are not in agreement with those obtained through comparing the
[\ion{O}{3}] $(\lambda4959+\lambda5007)/\lambda4363$ ratios and the hydrogen recombination spectra, suggesting that
the electron energy is unlikely to follow the $\kappa$-distributions over a global scale of the nebular regions.
Nevertheless, based on this observation alone, we cannot definitely rule out the presence of $\kappa$-distributed electrons  in some microstructures within 
nebulae.

\end{abstract}

%% Keywords should appear after the \end{abstract} command. 
%% See the online documentation for the full list of available subject
%% keywords and the rules for their use.

%% From the front matter, we move on to the body of the paper.
%% Sections are demarcated by \section and \subsection, respectively.
%% Observe the use of the LaTeX \label
%% command after the \subsection to give a symbolic KEY to the
%% subsection for cross-referencing in a \ref command.
%% You can use LaTeX's \ref and \label commands to keep track of
%% cross-references to sections, equations, tables, and figures.
%% That way, if you change the order of any elements, LaTeX will
%% automatically renumber them.
%%
%% We recommend that authors also use the natbib \citep
%% and \citet commands to identify citations.  The citations are
%% tied to the reference list via symbolic KEYs. The KEY corresponds
%% to the KEY in the \bibitem in the reference list below. 

\section{Introduction}

\label{sec:intro}

Plasma diagnostics using emission lines are of fundamental importance to understand the physical conditions and chemical compositions of photoionized gaseous nebulae such as planetary nebulae (PNe) and  \ion{H}{2} regions. 
A major problem in nebular physics, sometimes called the temperature and abundance discrepancy problem,
is that the electron temperatures and elemental abundances derived from collisionally excited lines (CELs) are significantly different from those derived from recombination spectra \citep[see, e.g.,][]{Peimbert1967,Liuetal.2000}. 
These discrepancies have motivated studies of new physical scenarios. For instance, 
\citet{Nemer2019} recently suggested that a process called Rydberg Enhanced Recombination,
which was never considered before, may contribute to the \ion{C}{2} and \ion{O}{2} recombination 
lines (RLs) in low-temperature photoionized plasma.
Among many proposals for solving the  temperature and abundance discrepancy problem, a rather disputable one is the presence of
nonthermal electrons \citep{Nicholls2012,Nicholls2013,Dopita2013}. In this scenario, the energy of free electrons
is assumed to follow a $\kappa$ function that has a superthermal tail in the otherwise 
Maxwell–Boltzmann (MB) energy distribution, while the usually assumed MB distribution is a special case of the $\kappa$ distribution in the limit of an infinite $\kappa$  index.
The intensities of CELs will be increased by the superthermal tail, leading to inappropriate results of the MB-based diagnostics.

The $\kappa$ distribution function have been commonly used to fit the
electron energy distribution in space plasma populated in the planetary magnetospheres,  heliospheres, solar wind, and solar corona \citep[e.g.][]{Vasyliunas1968, Feldmanetal.1975,Seely1987,Nicolaou2018}. It was found that
the $\kappa$ distribution can be theoretically deduced from the statistical mechanics 
of out-of-equilibrium system that are applicable for systems subject to long-range interactions
\citep{Livadiotis2009} . The origin of the $\kappa$ distribution has been a hot topic
for debate in the solar physics community \citep{Dudik2015,Testa2014, Nicolaou2019}.

However, the presence of $\kappa$-distributed electrons  is not supported by
the classical theory of photoionized gaseous nebulae. \citet{fer16} showed that the nonthermal
electrons in nebulae will be quickly relaxed to the MB energy distribution before they are able to
influence the excitation of CELs. The theoretical calculations of \citet{Draine2018} suggested that
the fraction of nonthermal electrons in PNe and \ion{H}{2} regions is too small to affect the
diagnostic results. These authors claimed that the temperature discrepancy problem must be caused by the spatial variations of electron temperatures. However, it should be noted that no theoretical model can
produce temperature variations to the level required to explain the observed temperature discrepancy \citep{stasinska2017}. From the theoretical perspective, it is fair to say that
the issue of the temperature variation hypothesis is not less pronounced
than that of the $\kappa$-distribution hypothesis.

Seldom efforts have been made to observationally determine the electron energy distribution in photoionized gaseous nebulae. 
\ion{C}{2} dielectronic RLs and \ion{H}{1} continuum emission spectra have been used to trace the electron energy distributions in PNe
\citep{storey2013,storey2014,Zhang2014}, but the reported results are inconclusive or even conflicting due to rather large uncertainties.
Under the assumption of $\kappa$ distributions, \citet{Zhang2016} estimated the $\kappa$ values for a sample of PNe and \ion{H}{2} regions by comparing the [\ion{O}{3}] $(\lambda4959+\lambda5007)/\lambda4363$ ratios and the \ion{H}{1} Balmer jump. There is no obvious correlation between the resultant $\kappa$ values and various physical properties of the nebulae, providing no support for the hypothesis that $\kappa$-distributions can be pumped by known physical mechanisms. 

Certainly, more observational evidences are needed to confirm or refute the $\kappa$-distribution hypothesis. If the $\kappa$-distributed electrons
are present throughout the nebula, a consistent $\kappa$ value would be
derived by comparing the intensity ratios of different diagnostic lines. 
Recently improved calculations of atomic data have confirmed that the MB temperatures derived from the \ion{O}{2}  RLs are remarkably lower than those derived from the [\ion{O}{3}] CELs \citep{Storey2017}. The purpose of the present paper is to investigate whether the \ion{O}{2}/[\ion{O}{3}] temperature discrepancy conforms with the $\kappa$-distribution hypothesis. 

Such an investigation usually requires the energy-tabulated rate coefficients, which are hardly ever available in the literature. 
Except for \citet{Storey20151,Storey20152}, which presented
the collision strengths for [\ion{O}{3}] CELs and the
recombination coefficients for hydrogen under the $\kappa$-distributions,
prior research mostly reported MB-integrated rate coefficients.
A similar situation has arisen in the study of solar plasma. In order to
investigate the atomic processes important for solar physics,
\citet{Hahn2015}  developed an easy method to approximately obtain the
 rate coefficients of $\kappa$-distribution plasma by summing up the
MB rate coefficients with appropriate weights. In this work we apply this method to investigate the \ion{O}{2} RLs  in nebulae with $\kappa$-distributed electrons.

This paper is structured as follows: Section~2 describes the methods used
to obtain the recombination coefficients of \ion{O}{2} RLs for $\kappa$ distributions. Section~3 presents
a uncertainty analysis and 
the $\kappa$ values derived from
the observed \ion{O}{2} RLs. In Section~4 we compare the results with those derived from \ion{H}{1} Balmer jump and discuss the implications. A summary are given in Section~5.

\section{Methodology} \label{sec:method}

The calculations of recombination rate coefficients are based on the energy-dependent recombination cross-sections integrating over the election energy distribution function $f$. The isotropic $\kappa$ distribution has the form
\begin{equation}\label{eqkappa}
f_\kappa(E,T_U,\kappa) = \frac{2}{\sqrt{\pi}}
\frac{\Gamma(\kappa+1)\sqrt{E}}{(\kappa-1.5)^\frac{3}{2}\Gamma(\kappa-0.5)}\left(\frac{1}{k_bT_U}\right)^{\frac{3}{2}}\left[{1+\frac{E}{(\kappa-1.5)k_bT_U}}\right]^{-\kappa-1},
\end{equation}
where $E$ is the electron energy,  $\Gamma$ is the Gamma function, and $k_b$ is the Boltzmann constant. The low-energy part of this function can be described by a MB function with a temperature of $T$, and the high-energy part resembles a power-law.
The unitless $\kappa$ index characterizes the deviation from the MB distribution with smaller value corresponding to a larger deviation,
and is always larger than 1.5.
In the limit of $\kappa\rightarrow\infty$, Equ.~(\ref{eqkappa}) reduces to the  MB function. 
The temperature $T_U$ characterizes the mean kinetic energy, and has the expression $T_U=\kappa T/(\kappa - 1.5)$.
Given the fact of $T_U$ larger than $T$,
 MB-based plasma diagnostics could result in higher CEL temperatures than RL temperatures as the high- and
 low-energy parts of $f$ more effectively contribute to the excitation of CELs and RLs, respectively.

\begin{figure*}
\centering
\plotone{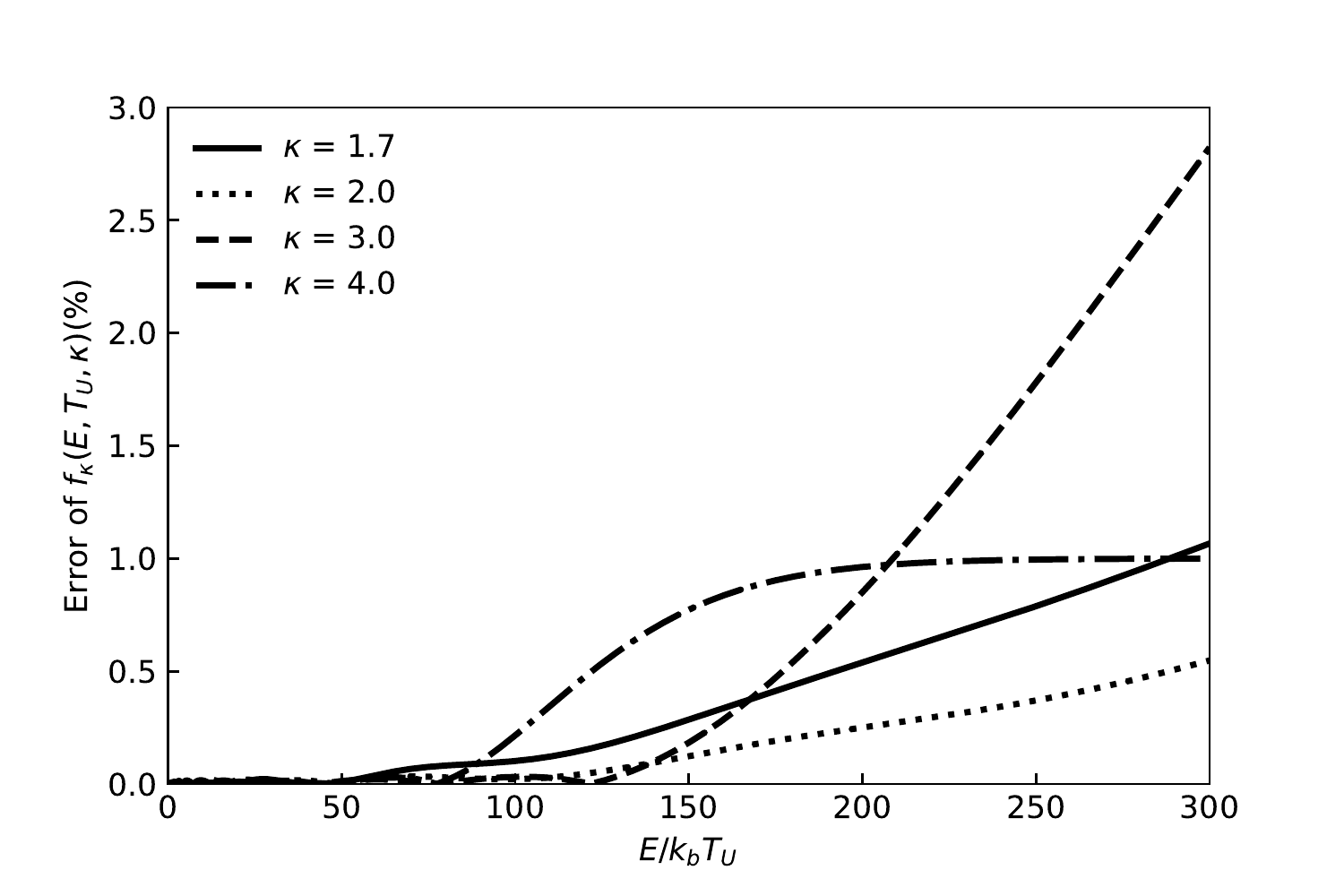}
%\gridline{\fig{error1.pdf}{0.8\textwidth}}
\caption{The errors of $f_{\kappa}(E,T_U,\kappa)$ derived through comparing
the results from Equs.~(\ref{eqkappa}) and (\ref{decomp}).}
\label{comparision}
\end{figure*}

The recombination rate coefficients take the form
\begin{equation}\label{rate}
\alpha(T)=\int \sigma (E)f(E,T)\sqrt{\frac{2E}{m_\mu}}dE,
\end{equation}
where $m_\mu$ is the reduced mass, and  $\sigma(E)$ is the cross section. 
Overwhelming published atomic data are for the plasma with a MB electron energy distribution $f_{\rm MB}(E,T)$.
In principle, the $\kappa$-distribution rate coefficients, $\alpha_\kappa(T_U,\kappa)$, can be accurately obtained by substituting $f_\kappa(E,T_U,\kappa)$ into Equ.~\ref{rate}. However, the tabulated cross sections are commonly unavailable. An approximate method was presented by
\citet{Hahn2015} to derive the $\kappa$-distribution rate coefficients from published MB rate coefficients $\alpha_{\rm MB}(T)$.
They found that $f_\kappa(E,T_U,\kappa)$  can be decomposed to a series of MB distribution functions, i.e.,
\begin{equation}\label{decomp}
f_{\kappa}(E,T_U,\kappa) = \sum_{j}c_j(\kappa)f_{\rm MB}(E,T_j)
\end{equation}
with $T_j=a_j(\kappa)T_U$, where the fitting parameters $a_j(\kappa)$ and $c_j(\kappa)$ are independent of $T_U$ and
satisfy  $\sum_{j}c_j(\kappa)=1$. 
The uncertainties associated with this method increase with increasing
$E/T_U$ and decreasing $\kappa$ index. However, as shown in Fig.~\ref{comparision}, even for the electron energy distribution with relatively low $\kappa$ indexes, the results are accurate within 3$\%$.
From Equ.~(\ref{rate}) and (\ref{decomp}), we have
\begin{equation}\label{alpha}
 \alpha_\kappa(T_U,\kappa)= \sum_{j}c_j(\kappa)\alpha_{\rm MB}(T_j),
\end{equation}
namely, the $\kappa$-dependent rate coefficients can be decomposed into several weighted MB rate coefficients.

 The purpose of \citet{Hahn2015} is to investigate the collision processes in
solar plasmas, such as the solar corona, which are generated by collisional ionization.
Dissimilarly, PNe and \ion{H}{2} regions are the photoionized gas characterized by 
much lower temperature and density,
where radiative decay predominates over collisional decay for hydrogen. 
Nevertheless, the decomposition approach does not depend on the specific physical 
conditions and atomic processes, and applies to the calculation of any atomic data that can be expressed by Formula~(\ref{rate}) (e.g., recombination coefficients and collision strengths). 
In low-density nebulae, because the effect of collisional transitions is likely negligible, captures and downward-radiative transitions are the only processes to 
produce RLs, and thus the effective recombination coefficient to the $i$-th level
is a linear weighted sum of the recombination coefficients to the 
levels of $\ge i$. As a result, the approximate method can be employed to compute
the effective recombination coefficients (and thus the emissivities) of RLs from the plasma with $\kappa$-distributions. In this work we focus on the commonly detected
\ion{H}{1} and \ion{O}{2} RLs in photoionized gaseous nebulae.

\section{Results} \label{sec:results}
%\subsection{Uncertainty analysis based on the \ion{H}{1} RLs}
\subsection{Viability of the decomposition approach in determining the emission coefficients
of  H~{\sc i} RLs}
\label{sec:H}

Taking the $a_j(\kappa)$ and $c_j(\kappa)$ parameters provided in \citet{Hahn2015}, we computed the
$\kappa$-dependent emission coefficients $\epsilon(N_{\rm e},T_U,\kappa)$
for the hydrogen Balmer and Paschen lines using the  decomposition approach by replacing $\alpha_\kappa(T_U,\kappa)$ and $\alpha_{\rm MB}(T_j)$ in Equ.~(\ref{alpha}) with $\epsilon(N_{\rm e},T_U,\kappa)$ and $\epsilon(N_{\rm e},T_j,\infty)$.
\citet{Storey20152} reported $\epsilon(N_{\rm e},T_U,\kappa)$ of \ion{H}{1} lines obtained 
from first-principle calculations,  allowing us to validate the results of this approximate approach.  
When the $\kappa$ index is sufficient large, $\epsilon(N_{\rm e},T_j,\kappa)$ is practically the same
with the corresponding MB value. Thus in the calculations with Equ.~(\ref{alpha}) we have used   $\epsilon(N_{\rm e},T_j,10^6)$ given by \citet{Storey20152} as 
a substitute of the MB emission coefficients. In summing the weighted
MB emission coefficients, we have ignored the terms of $T_j>10^5K$ in that the recombination is essentially
insignificant at such high temperatures.

\begin{figure*}
\plotone{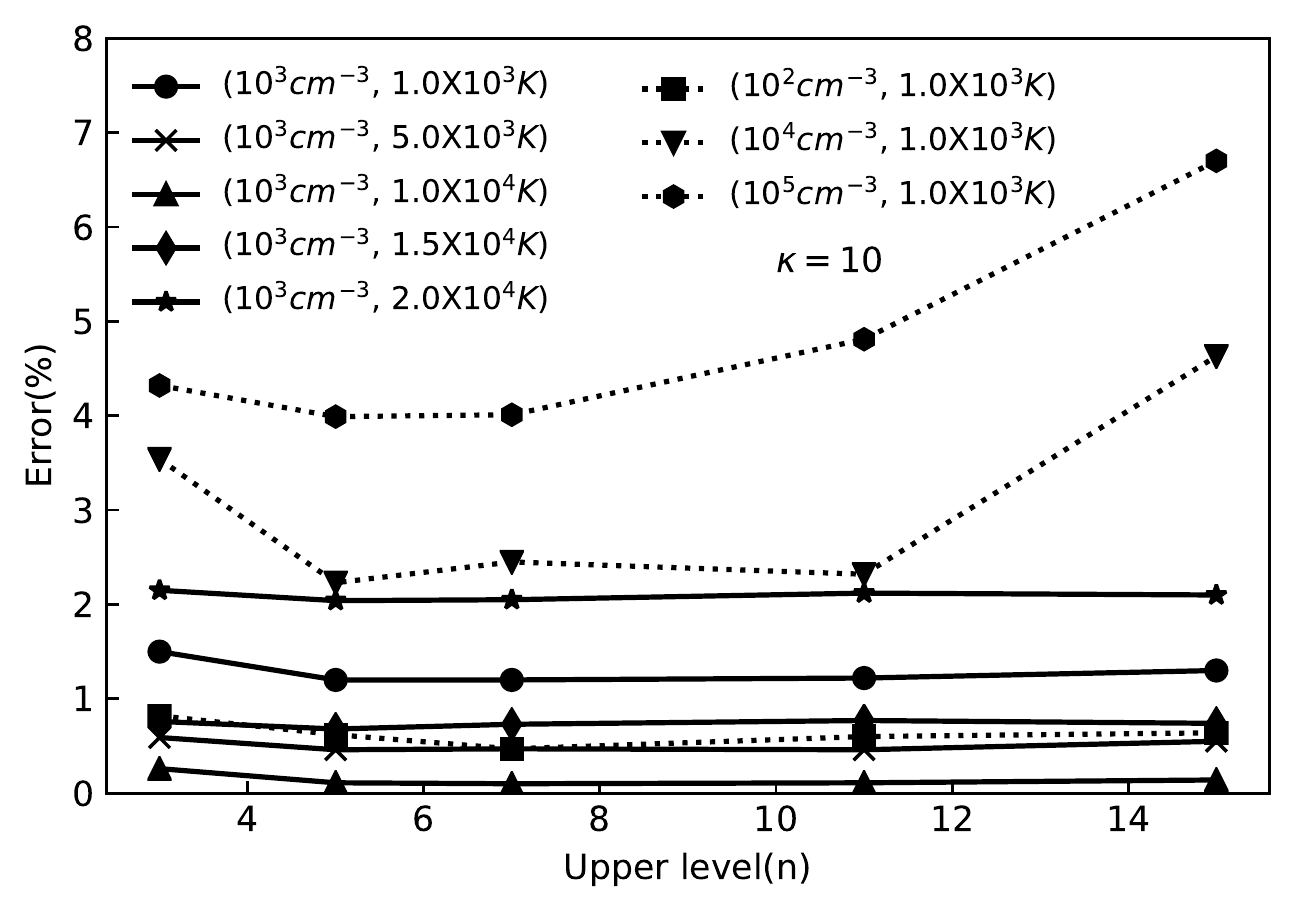}
\caption{The errors of $\epsilon(N_{\rm e},T_U,\kappa)$ at $\kappa=10$ for the \ion{H}{1} Balmer decrements ($n$--$2$) obtained through the decomposition approach. Various $N_{\rm e}$ and $T_U$ values have been assumed.}
\label{comparisionh1a}
\end{figure*}

\begin{figure*}
\plotone{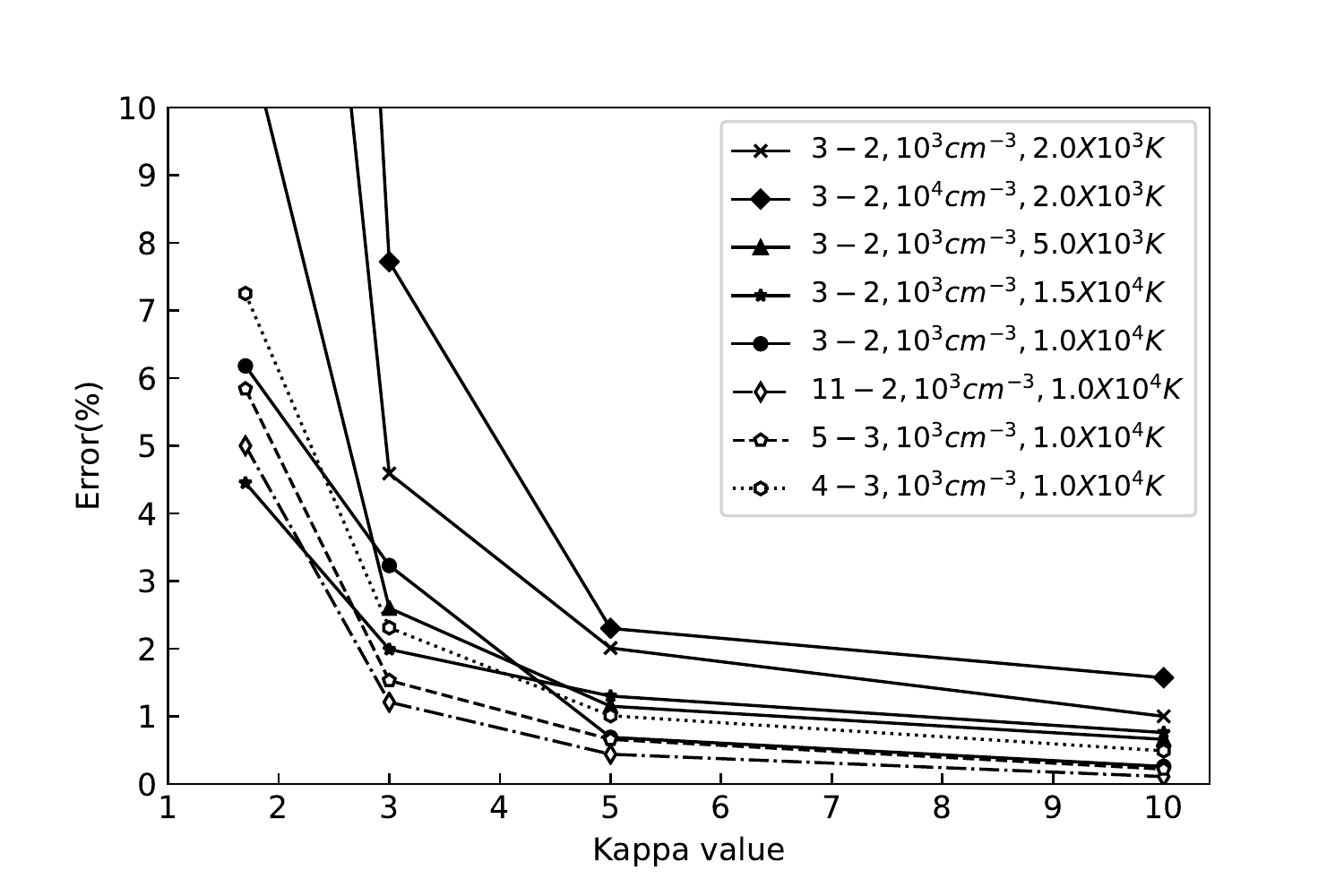}
\caption{ The errors of $\epsilon(N_{\rm e},T_U,\kappa)$ of a few  \ion{H}{1} RLs obtained
through the decomposition approach as a function of $\kappa$. The upper and lower levels of the  \ion{H}{1} transitions and the assumed temperatures and densities are shown in the up-right position.}
\label{comparisionh1b}
\end{figure*}

The uncertainties of  $\epsilon(N_{\rm e},T_U,\kappa)$ can be estimated through comparing our results with
those given by \citet{Storey20152}, as illustrated in Figs.~\ref{comparisionh1a} \& \ref{comparisionh1b}.
An inspection of Fig.~\ref{comparisionh1a} shows that the errors increase with increasing electron 
density, suggesting that the impact of collisions is becoming increasingly important and
cannot be ignored in very dense conditions. Nevertheless, for the typical physical conditions of PNe
and \ion{H}{2} regions ($N_{\rm e}<10^4$\,cm$^{-3}$ and $T_U<2\times10^4$\,K), the errors are up to 
5$\%$.  Fig.~\ref{comparisionh1b} presents the errors of $\epsilon(N_{\rm e},T_U,\kappa)$ for a few Balmer 
and Paschen lines. The errors tend to decrease for the lines with higher upper levels, and dramatically increase
with decreasing $\kappa$ values. At the range of  $\kappa>5$, the errors are typically less than 3$\%$. For extreme 
 $\kappa$-distribution ($\kappa<3$), it could be up to 10$\%$. 
  %For a given temperature and density, the results of the \ion{H}{1} 3--2 transition (H$\alpha$) generally have larger errors compared to the others. 
 Fig.~\ref{comparisionh1b} also plots the errors for the \ion{H}{1} 3--2 line at various 
 temperatures and densities, from which we can clearly see that even under extreme conditions
 the results are relatively reliable at the range of $\kappa>3$. 
 Consequently, the uncertainty analysis based on 
\ion{H}{1} RLs strongly suggests that in typical nebular conditions, the decomposition approach can provide a close approximation to investigate the $\kappa$-dependent emissivities of RLs. But one should take caution when
applying this method to the plasma with extremely low $\kappa$ index.

\subsection{The Kappa index derived from the O~{\sc ii} RL ratios}\label{sec:O II}

\begin{figure}
\plotone{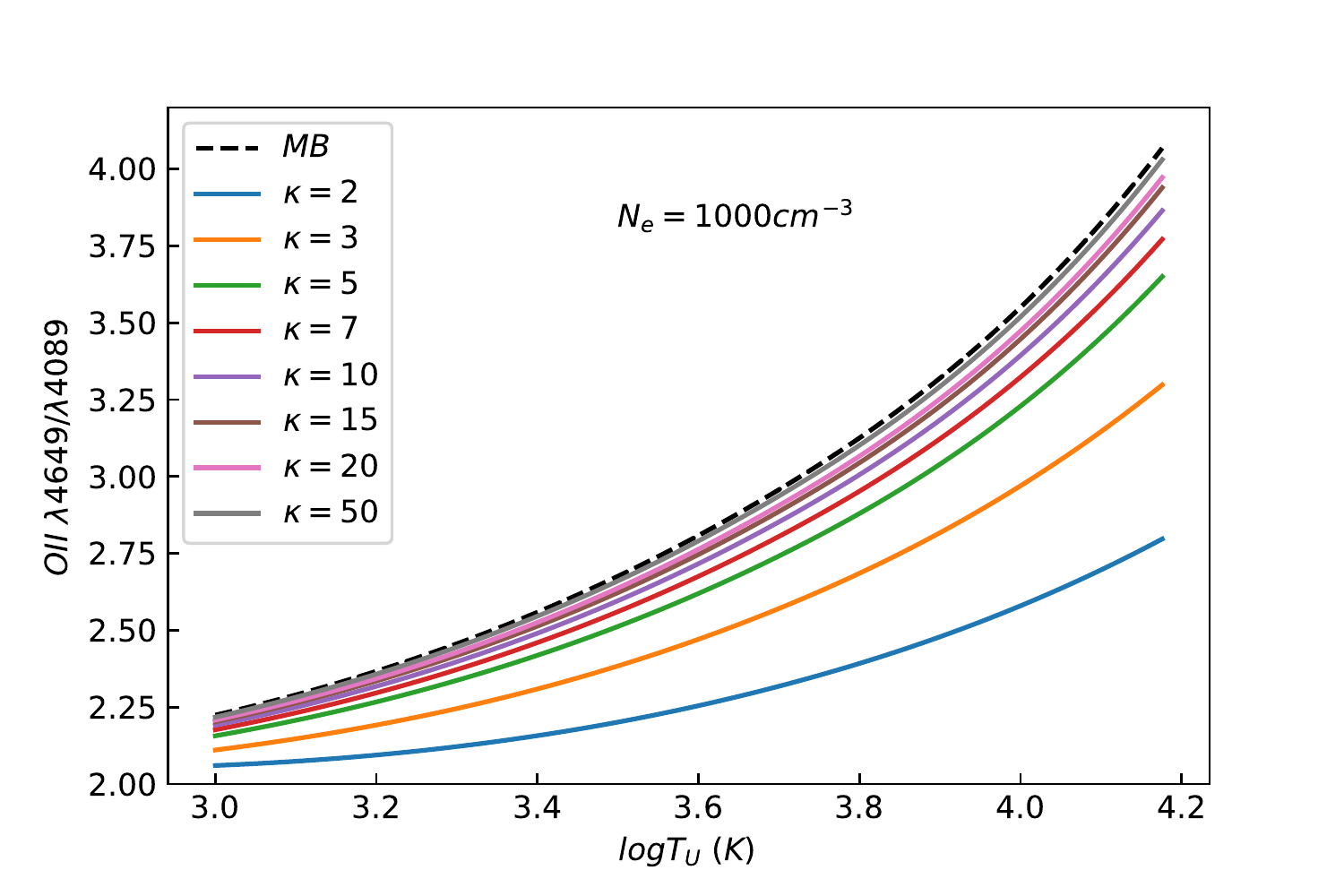}
\caption{The theoretical \ion{O}{2} $\lambda4649$/$\lambda4089$ intensity ratio as a function of $T_U$  for various
$\kappa$ indexes. $N_{\rm e}$ is assumed to be $10^3$\,cm$^{-3}$. \label{figoii}}
\end{figure}

In order to investigate the influence of $\kappa$-distributed electrons on the temperature diagnostics of RLs, we computed the $\kappa$-dependent emissivities of \ion{O}{2} RLs using the decomposition approach. 
In the calculations, we have set a constant density of $10^3$\,cm$^{-3}$ and
taken the  MB recombination coefficients of \ion{O}{2} lines in case B recently reported by \citet{Storey2017}. 
The  \ion{O}{2} $\lambda4649$/$\lambda4089$ intensity ratio has been commonly used to determine the electron temperatures of PNe \citep{Wesson2003,Fang2013,McNabb2016}. 
As stated by \citet{Storey2017}, the \ion{O}{2} $\lambda\lambda4649,4089$ lines originate from high-$J$ levels, and thus both strongly depend on the population of the $^3$P$_2$ 
level of O$^{2+}$. As a result,  their intensity ratio is rather insensitive to density, and can serve as a good temperature indicator. 
In Table~\ref{OIIratiodata}, we present the theoretical \ion{O}{2} $\lambda4649$/$\lambda4089$ intensity ratio for different $\kappa$ and  $T_U$ values,
which can be readily used to determine the physical
conditions of $\kappa$ distribution plasma.
As illustrated in Fig.~\ref{figoii}, the \ion{O}{2} $\lambda4649$/$\lambda4089$ ratio increases with increasing $\kappa$ and  $T_U$. It is apparent that this ratio is
more sensitive to $\kappa$  in high-temperature and low-$\kappa$ ranges,  and is virtually 
indistinguishable between MB and  $\kappa$ distributions in the range of  $\kappa>20$. Therefore, 
when we use the \ion{O}{2} $\lambda4649$/$\lambda4089$ ratio to determine the $\kappa$ index,
the uncertainties dramatically increase with increasing $\kappa$.

\begin{figure}
\plotone{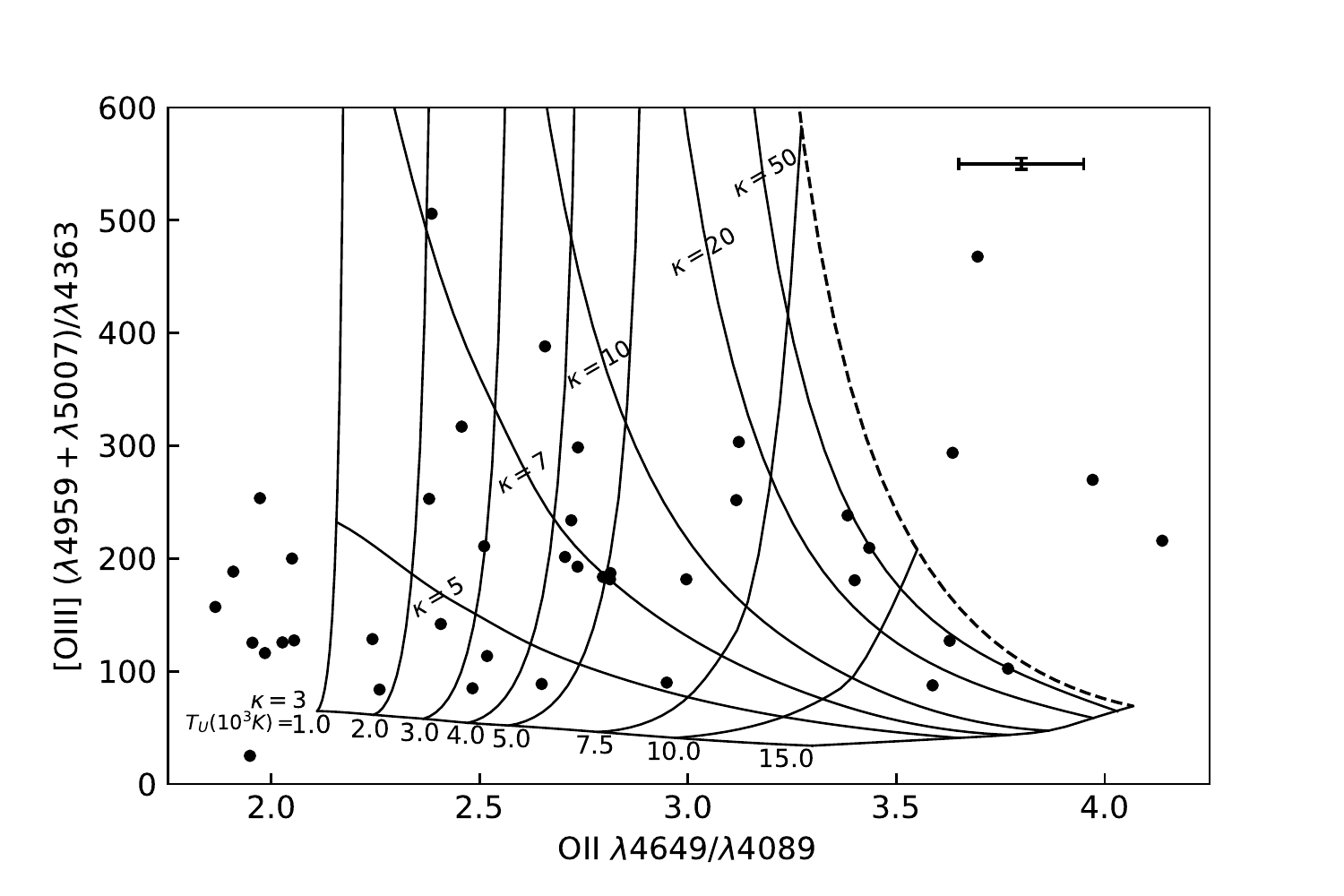}
\caption{The [\ion{O}{3}] CEL ratios versus the \ion{O}{2} RL ratios. The solid curves are the theoretical intensity ratios as functions of $\kappa$ and $T_U$, which decay to the 
MB one (dashed curve) in the limit of $\kappa\rightarrow\infty$.
The filled circles represent the observed values with a typical error bar marked
in the up-right corner. \label{figdia}}
\end{figure}

In a homogeneous nebula, the [\ion{O}{3}] CELs at 4959, 5007, and 4363\,{\AA} arise from the same region with the \ion{O}{2} RLs.
\citet{Storey20151} reported the effective collision strengths for excitation and
de-excitation of [\ion{O}{3}] lines with $\kappa$-distributed electron energies, which
can be used to derive the theoretical [\ion{O}{3}] $(\lambda4959+\lambda5007)/\lambda4363$ 
intensity ratio as functions of $\kappa$ and $T_U$. Then $\kappa$ and $T_U$  can be simultaneously derived by comparing the observed 
[\ion{O}{3}] $(\lambda4959+\lambda5007)/\lambda4363$ and \ion{O}{2} $\lambda4649$/$\lambda4089$ ratios. Using the observational data from the literature,
we employed this method to derive the $\kappa$ values for a sample of PNe  (Table~\ref{Tresults}). The results are illustrated in Fig.~\ref{figdia}. The uncertainties of the \ion{O}{2} ratio were evaluated to be around 10--20\,$\%$. 
From Fig.~\ref{figdia} we can see that although the \ion{O}{2} ratio has a much larger uncertainty than the [\ion{O}{3}] ratio, it does not introduce large uncertainty in determining $\kappa$
when the $\kappa$ index is very low ($<5$).
No data point lies on the theoretical curve corresponding to the MB
energy distribution with most lying on the down-left side.
Under the framework of MB energy distribution, the [\ion{O}{3}] $(\lambda4959+\lambda5007)/\lambda4363$ ratios suggest electron temperatures of 
7500--15000\,K, while significantly lower temperatures are obtained from
the \ion{O}{2} $\lambda4649$/$\lambda4089$ ratios (500--15000\,K), according with previous results reported
by other authors.
As shown in  Fig.~\ref{figdia},
the temperature discrepancy can be largely eliminated by invoking the
$\kappa$ distributions. A few objects exhibit extremely low  \ion{O}{2} $\lambda4649$/$\lambda4089$ ratios,
which may be partially attributed to the blending of the \ion{O}{2} $\lambda4089$ line with a \ion{S}{4} line at
4088.86\,{\AA}, as suggested by \citet{pei13}.

As shown in  Table~\ref{Tresults}, most of the PNe have a $\kappa$ value of $<10$, indicating a large deviation
from the MB electron energy distribution.
For comparison, Table~\ref{Tresults} gives the  $\kappa$ and $T_U$ derived by \citet{Zhang2016}
through comparing [\ion{O}{3}] $(\lambda4959+\lambda5007)/\lambda4363$ ratio and the  \ion{H}{1} Balmer jump.
The values obtained by the two methods are in poor agreement.
 Table~\ref{Tresults} also lists the most recently published abundance discrepancy factor (ADF) for O$^{2+}$,
 which is defined as the ratio between the O$^{2+}$ abundances derived from \ion{O}{2} RLs and [\ion{O}{3}] CELs.
No significant correlation could be found between the $\kappa$ index and the ADF.
Despite this, the PNe with extremely large ADF, such as Hf~2-2 \citep{liu06}, exhibit relatively
lower $\kappa$ values.

\section{Discussion}

\begin{figure}
\plotone{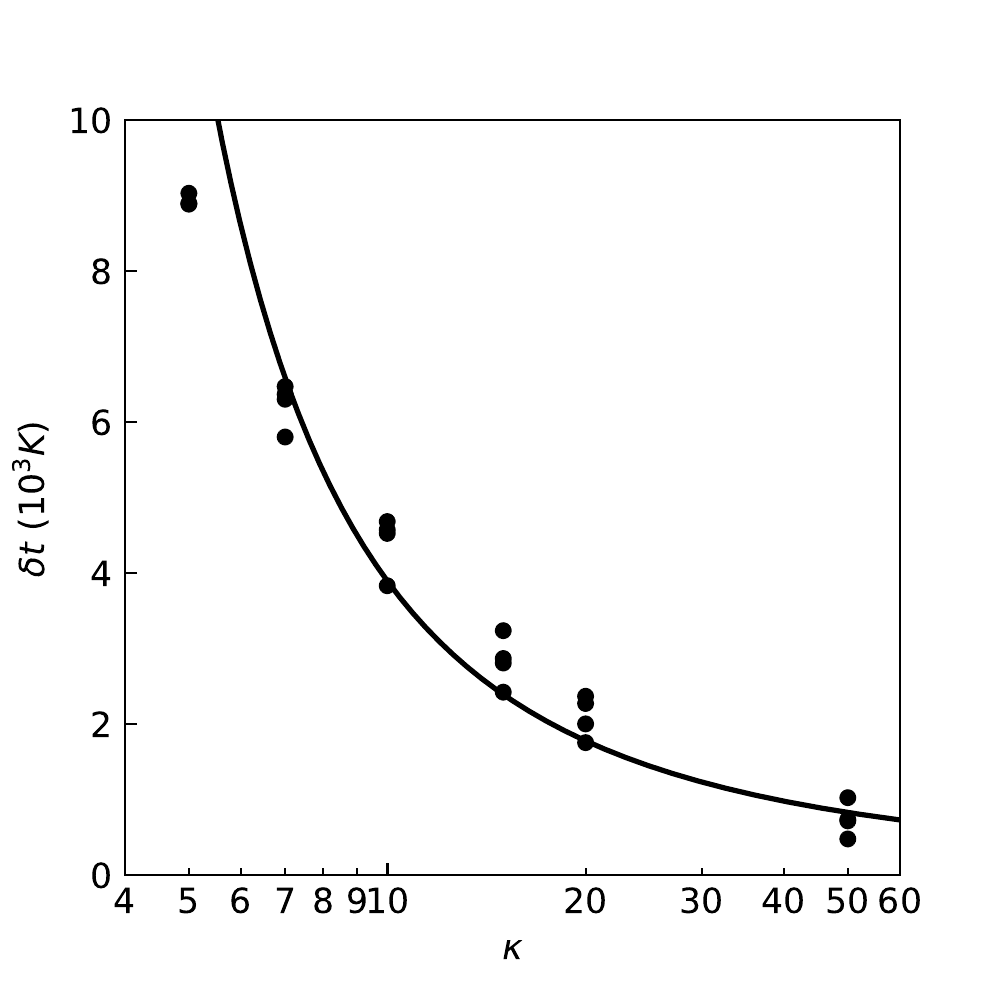}
\caption{$\delta t$ versus $\kappa$. The filled circles are
the theoretical $\delta t$ values calculated by fixing $T_{\rm MB}$([\ion{O}{3}])$=8000$, 10000, 12000, and 140000\,K. The solid curve represents an empirical fitting function. \label{fits}}
\end{figure}

In the present study, we investigated the effect of $\kappa$-distributed elections on the emissivities of
\ion{H}{1} and \ion{O}{2} RLs as well as on the plsama diagnostics based on RLs 
by using a decomposition approach. In typical nebular conditions, the 
emissivities of \ion{H}{1} RLs obtained by this approximate method are in excellent agreement with the ab initio calculations, demonstrating the viability of this method. We found that the presence of
$\kappa$-distributed elections can reduce the \ion{O}{2} $\lambda4649$/$\lambda4089$ intensity ratio, and
hence lead to significantly underestimated electron temperatures for MB distributions.
\citet{McNabb2016} used the \ion{O}{2} $\lambda4649$/$\lambda4089$ ratio to determine the electron temperatures
of a PN sample, which are about 3000\,K on average, while the electron temperatures derived from the [\ion{O}{3}] 
CELs are typically about 10000\,K. Our results indicate that the [\ion{O}{3}]  and  \ion{O}{2}  temperatures
can reach a common value through tuning the $\kappa$ index. The $\kappa$-distribution hypothesis, therefore, seems
to provide a plausible explanation for the CEL/RL temperature discrepancy problem. As
shown in Fig.~\ref{fits}, the relationship between the $\kappa$ index and the [\ion{O}{3}]/\ion{O}{2} MB-temperature difference $\delta t$
can be  approximated with an empirical function 
\begin{equation}\label{eqfit}
    \kappa=1.5\exp{\frac{3.25}{\delta t^{0.40}}},
\end{equation}
where $\delta t$ euqals to $T_{\rm MB}$([\ion{O}{3}])$-T_{\rm MB}$(\ion{O}{2}) in unit of $10^3$\,K. This formula can be used to estimate the $\kappa$ values 
from the previously reported \ion{O}{2} temperatures.

\begin{figure}
\plotone{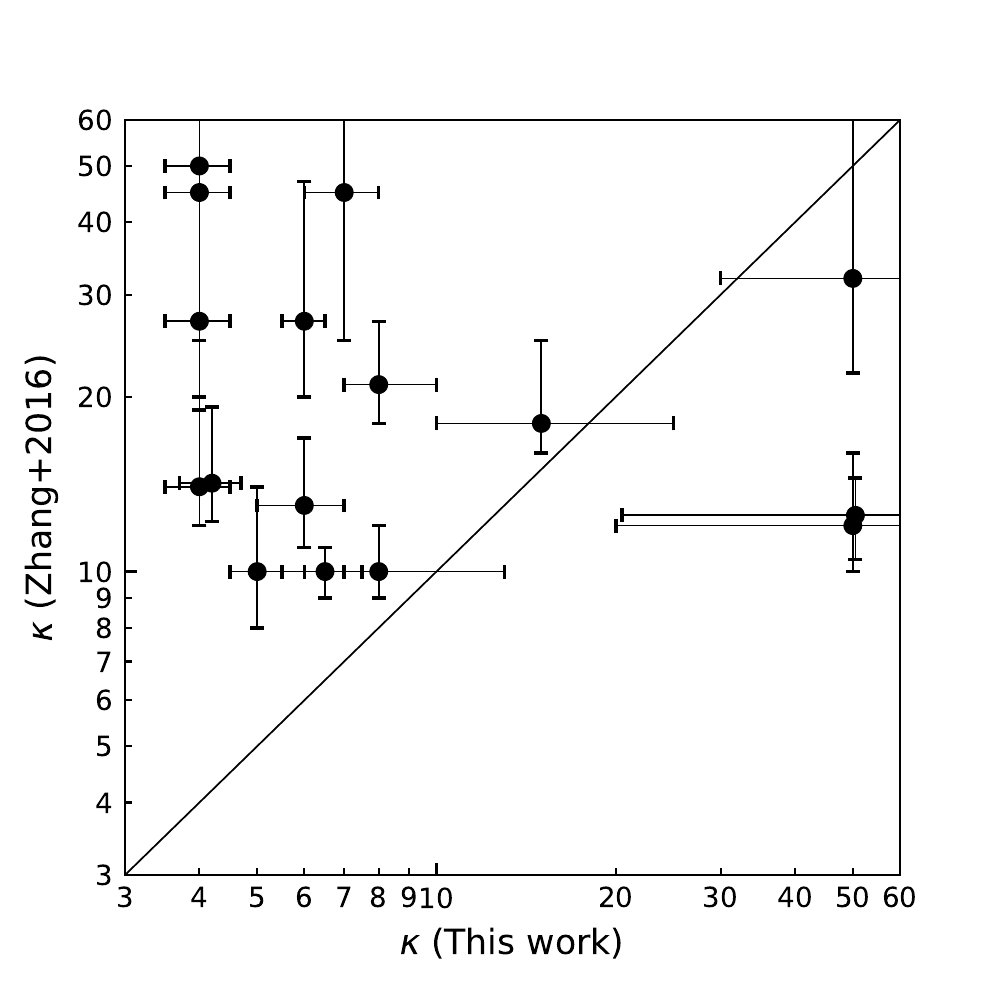}
\caption{Comparison between the  $\kappa$ indexes derived from \ion{O}{2} RLs (this work) and Balmer jump \citep{Zhang2016}. The solid line is a $y=x$ plot. \label{figcomp1}}
\end{figure}

\begin{figure}
\plotone{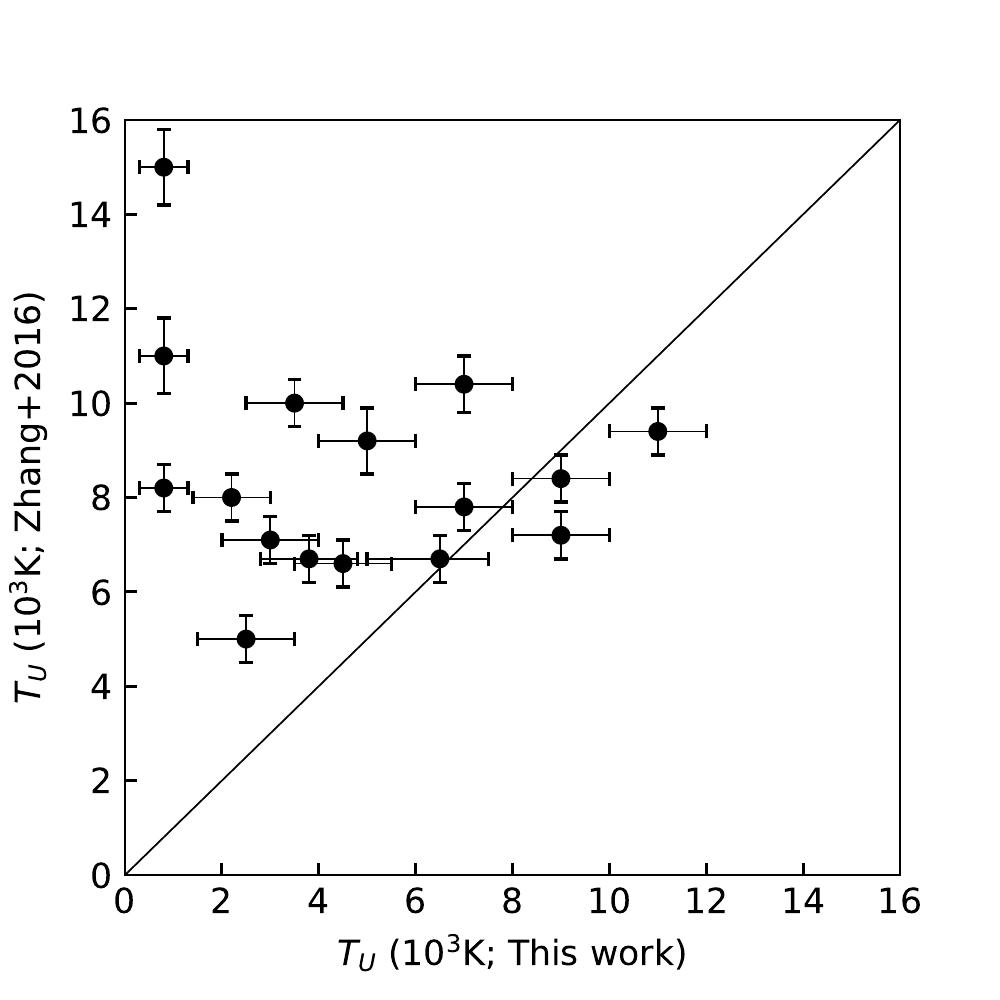}
\caption{Comparison between the $T_U$ values derived from \ion{O}{2} RLs (this work) and Balmer jump \citep{Zhang2016}. The solid line is a $y=x$ plot. \label{figcomp2}}
\end{figure}

However, contrary to the expectation of the  $\kappa$-distribution hypothesis, the $\kappa$ values derived
from the  [\ion{O}{3}] CELs and  \ion{O}{2} RLs are not consistent with those derived 
by \citet{Zhang2016} from the  [\ion{O}{3}] CELs and   \ion{H}{1} Balmer jump. 
The latter showed that the average $\kappa$ values of PNe are about 30, much larger than the current results.
The $\kappa$ and $T_U$ values derived by us and those by \citet{Zhang2016} are compared in Figs.~\ref{figcomp1} and \ref{figcomp2}, respectively. If the superthermal electrons are homogeneously distributed over the whole nebula, a consistent $\kappa$ value should be derived from different diagnostics. 
However, from Figs.~\ref{figcomp1} and \ref{figcomp2}, we can see that $\kappa$ and $T_U$ derived from the  \ion{O}{2} ratio are 
systematically  smaller than those derived from the \ion{H}{1} Balmer jump.
This is a strong evidence against the presence of globally distributed
superthermal electrons, and thus is compatible with current theoretical propositions that
no known physical mechanism can pump a $\kappa$ electron energy distribution over the whole photoionized gaseous nebula \citep{fer16,Draine2018}.

A hypothesis that cannot be ruled out is that $\kappa$-distributed electrons perhaps exist only at a small spatial scale. 
\citet{Zhang2016} obtained a similar fitting function for $\kappa$ versus 
$T_{\rm MB}$([\ion{O}{3}])$-T_{\rm MB}$(\ion{H}{1}). Comparing their formula with
Equ.~(\ref{eqfit}), we can conclude that $T_{\rm MB}$(\ion{O}{2})  more sensitively depends on $\kappa$ than does $T_{\rm MB}$(\ion{H}{1}). It follows that
if there was $\kappa$-distribution plasma embedded within the `normal' nebula, 
the \ion{O}{2} $\lambda4649$/$\lambda4089$ ratio would be affected significantly more
than the \ion{H}{1} Balmer jump. This provides a possible explanation for the smaller
 $\kappa$ values obtained by us. 
 If this is the case, the  $\kappa$ indices listed in Table~\ref{Tresults} are 
 only upper limits and, in principle, a comparison of two pairs of  \ion{O}{2} RLs can
 impose a more stringent constraint on the 
 $\kappa$-index of the non-thermal component. To develop a better understanding of nebular physical conditions, we need to construct models comprising adjustable parameters representing the properties of the non-thermal component (temperature, density, $\kappa$-index, filling factor, etc.) to match spectroscopic observations in particular of RLs. For this purpose, the computations of $\kappa$-based atomic data for 
 more RLs from O$^+$ and other ions are required, which are beyond the scope of this paper.

 It is well established that the 
 $\kappa$-distribution plasma in the solar system is mostly confined in local regions.
 If such plasma could be preserved until the PN phase or the same pumping
 mechanism works in photoionized gaseous nebulae, the plasma diagnostics of PNe would be greatly influenced.  
Although this is admittedly a highly speculative conjecture,
 further theoretical studies are desirable.

\section{Conclusion} \label{sec:Summary}

In this work we computed the $\kappa$-dependent emissivities of RLs using a decomposition method, aiming to find observational evidence supporting or
rejecting the postulated presence of $\kappa$-distributed electrons in PNe and 
\ion{H}{2} regions.
 The validity of this approximate method was verified through an analysis of
\ion{H}{1} RLs,  which suggests that reliable results can be obtained in the
range of $\kappa > 3$. 
We showed that the \ion{O}{2} $\lambda4649$/$\lambda4089$ intensity ratio 
has a moderate dependence on the $\kappa$ index in the range of $\kappa < 20$, and the introduction 
of $\kappa$ electron energy distribution allows us to reconcile the
incongruent MB temperatures derived from [\ion{O}{3}] CELs and \ion{O}{2} RLs.
However, the $\kappa$  values estimated from the [\ion{O}{3}]/\ion{O}{2} 
temperature discrepancies are generally lower than those obtained from
the [\ion{O}{3}]/\ion{H}{1} ones, disfavoring the global
presence of $\kappa$-distributed electrons over the whole nebulae. Therefore, we can conclude that
if the $\kappa$-distribution hypothesis holds, the superthemral electrons must
be distributed within small-scale regions.
Further investigation of $\kappa$-dependent emissivities of other RLs such as \ion{He}{1} is strongly
recommended.

\acknowledgments
We are grateful to the anonymous referee for constructive comments that contributed to improve the manuscript.
This work was supported by National Science Foundation of China (NSFC, Grant No. 11973099).

\bibliography{references}{}
\bibliographystyle{aasjournal}

\clearpage

\clearpage

\begin{table}
    \centering
    \caption{The computed \ion{O}{2} $\lambda4649$/$\lambda4089$ intensity ratios \label{OIIratiodata}}
\begin{tabular}{c|cccccccccc}
\hline
\diagbox[width=11em]{$T_U(K)$}{$\kappa$} & 2 & 3 & 5  & 7 & 10 & 15 &20 &50 & MB \\
\hline
1000  & 2.0601 & 2.1106 & 2.1566 & 2.1764 & 2.1912 & 2.2023 & 2.2079 & 2.2175 & 2.2241 \\
2000  & 2.1221 & 2.2460 & 2.3378 & 2.3730 & 2.3986 & 2.4188 & 2.4289 & 2.4461 & 2.4571 \\
5000  & 2.3172 & 2.5696 & 2.7400 & 2.8037 & 2.8508 & 2.8848 & 2.9041 & 2.9350 & 2.9556 \\
7500  & 2.4586 & 2.7843 & 3.0003 & 3.0804 & 3.1401 & 3.1830 & 3.2067 & 3.2457 & 3.2719 \\
10000  & 2.5794 & 2.9697 & 3.2288 & 3.3231 & 3.3936 & 3.4467 & 3.4730 & 3.5191 & 3.5499 \\
12500  & 2.6891 & 3.1371 & 3.4404 & 3.5481 & 3.6293 & 3.6931 & 3.7220 & 3.7744 & 3.8089 \\
15000 & 2.7969 & 3.2982 & 3.6500 & 3.7715 & 3.8647 & 3.9394 & 3.9725 & 4.0306 & 4.0682 \\
\hline
\end{tabular}
\end{table}

%\clearpage

\startlongtable
\begin{deluxetable}{@{\extracolsep{4pt}}lccccrc@{}}
%\tablenum{2}
\tablecaption{The resultant $T_U$ and $\kappa$ values of PNe  \label{Tresults}}
%\tablewidth{1pt}
\tablehead{
\colhead{Object} & \multicolumn{2}{c}{$T_U(K)$}  & \multicolumn{2}{c}{$\kappa$} 
 & \colhead{ADF}  & \colhead{Ref.}
 \\  \cline{2-3}  \cline{4-5}
%  \\ \cmidrule(r){2-3}  \cmidrule(r){4-5}
\colhead{} & \colhead{Zhang+2016} & \colhead{This work} & \colhead{Zhang+2016} &
 \colhead{This work} & \colhead{} & \colhead{}
}
\startdata
Cn 2-1    & ...                  & $5000^{+1500}_{-1000}$    &...               & $7^{+1}_{-1}$         & 1.02   & 6  \\
H 1-35    & ...                  & $<1000$                   & ...              & $5^{+0.5}_{-0.5}$     & 1.04   & 6  \\
H 1-42    & ...                  & $4500^{+1000}_{-1000}$    &$>60$             & $6.5^{+1}_{-1}$       & 1.04   & 6\\
H 1-50    & ...                  & $2500^{+1000}_{-1000}$    & ...              & $4.5^{+0.5}_{-0.5}$   & 1.05   & 11 \\
Hf 2-2     & ...                  & $1800^{+1000}_{-1000}$    & ...              & $3.5^{+0.5}_{-0.5}$   & 83.00  & 9 \\
Hu 2-1    &$9000^{+800}_{-8000}$ & ...                       & $50^{+\infty}_{-20}$    & ...            & 4.00   & 4\\
Hen 2-73  & ...                  & $<1000$                   & ...              & $4^{+0.5}_{-0.5}$     & 2.29   & 11 \\
IC 4191   & $9200^{+700}_{-700}$ & $5000^{+1000}_{-1000}$    & $45^{+\infty}_{-20}$    & $7^{+1}_{-1}$  & 2.40          & 3 \\
IC 4699   & ...                  & $<1000$                   & ...              & $4^{+0.5}_{-0.5}$     & 1.09   & 6 \\
IC 4846   & $8200^{+500}_{-500}$ & $<1000$                   & $14^{+5}_{-2}$   & $4^{+0.5}_{-0.5}$     & 2.91   & 4 \\
IC 5217   & ...                  & $<1000$                   & ...              & $4^{+0.5}_{-0.5}$     & 2.26   & 4 \\
M 1-33    & ...                  & $4200^{+800}_{-800}$      & ...              & $8.5^{+1.5}_{-1.5}$   & 2.33   & 11 \\
M 1-42    & ...                  & $1500^{+1000}_{-1000}$    & ...              & $4^{+0.5}_{-0.5}$     & 22.00  & 9 \\
M 1-60    & ...                  & $7000^{+1000}_{-1000}$    & ...              & $18^{+30}_{-5}$       & 2.75   & 11\\
M 2-23    & ...                  & $3500^{+1000}_{-1000}$    & ...              & $4.5^{+0.5}_{-0.5}$   & 1.40   & 6 \\
M 2-31    & ...                  & $<1000$                   & ...              & $4^{+0.5}_{-0.5}$     & 2.42   & 11 \\
M 2-36    & $6700^{+500}_{-500}$ & $3800^{+1000}_{-1000}$    & $21^{+6}_{-3}$   & $8^{+2}_{-1}$         & 6.90   & 1 \\
M 2-39    & ...                  & $<1000$                   & ...              & $<3$                  & 0.40   & 6 \\
M 3-21    & ...                  & $4200^{+1000}_{-1000}$    & ...              & $6.5^{+1}_{-1}$       & 1.05   & 6 \\
M 3-32    & $5000^{+500}_{-500}$ & $2500^{+1000}_{-1000}$    & $10^{+1}_{-1}$   & $6.5^{+1}_{-1}$       & 17.75  & 6 \\
M 3-33    & $6700^{+500}_{-500}$ & $6500^{+1500}_{-1000}$    & $10^{+2}_{-1}$   & $8^{+5}_{-1}$         & 6.56   & 6 \\
M 3-34    & $9400^{+500}_{-500}$ & $11000^{+1000}_{-1000}$   & $12^{+4}_{-2}$   & $50^{+\infty}_{-30}$  & 4.23   & 4 \\
Me 2-2    & ...                  & $9000^{+1000}_{-1000}$    & $>60$            & $30^{+\infty}_{-15}$  & 2.10   & 4 \\
NGC 3242  &$15000^{+1000}_{-1000}$ & $<1000$                 & $27^{+\infty}_{-15}$   & $4^{+0.5}_{-0.5}$       & 2.20 & 3 \\
NGC 3918  & ...                  & $12500^{+1000}_{-1000}$   & $>60$            & $50^{+\infty}_{-30}$  & 2.30   & 3 \\
NGC 5307  &$11000^{+800}_{-800}$ & $<1000$                   & $40^{+\infty}_{-25}$   & $4^{+0.5}_{-0.5}$       & 1.95 & 2 \\
NGC 5315  & $8700^{+600}_{-600}$ & ...                       & $60^{+\infty}_{-30}$   & ...             & 2.00   & 10 \\
NGC 5882  & $8000^{+500}_{-500}$ & $2200^{+800}_{-800}$      & $27^{+20}_{-7}$  & $6^{+0.5}_{-0.5}$     & 2.10   & 3 \\
NGC 6153  & $6600^{+500}_{-500}$ & $4500^{+1000}_{-1000}$    & $14^{+5}_{-2}$   & $4^{+0.5}_{-0.5}$     & 9.20   & 9 \\
NGC 6210  & ...                  & $4200^{+1000}_{-1000}$    & $>60$            & $7^{+1}_{-1}$         & 3.10   & 8\\
NGC 6439  & ...                  & $5000^{+1000}_{-1000}$    & $>60$            & $7^{+2}_{-0.5}$       & 6.16   & 6 \\
NGC 6620  & $8400^{+500}_{-500}$ & $9000^{+1000}_{-1000}$    & $32^{30}_{-10}$  & $50^{+\infty}_{-20}$  & 3.19   & 6 \\
NGC 6803  & $7800^{+500}_{-500}$ & $7000^{+1000}_{-1000}$    & $18^{+7}_{-2}$   & $15^{+10}_{-5}$       & 2.71   & 4 \\
NGC 6807  &$10000^{+500}_{-500}$ & $3500^{+1000}_{-1000}$    & $50^{+\infty}_{-25}$  & $4^{+0.5}_{-0.5}$    & 2.00    & 4 \\
NGC 7009  & $7100^{+500}_{-500}$ & $3000^{+1000}_{-1000}$    & $13^{+4}_{-2}$   & $6^{+0.5}_{-0.5}$     & 5.00   & 7 \\
NGC 7026  & $7800^{+500}_{-500}$ & ...                       & $25^{+12}_{-7}$  & ...                   & 3.36   & 4 \\
NGC 7027  & ...                  & $12000^{+1500}_{-1500}$   & $>60$            & $15^{+5}_{-5}$        & 1.29   & 5 \\
Vy 1-2    & $7200^{+500}_{-500}$ & $9000^{+1500}_{-1000}$    & $12^{+2}_{-2}$   & $50^{+\infty}_{-30}$  & 6.17   & 5 \\
Vy 2-1    & ...                  & $2000^{+1000}_{-1000}$    & ...              & $7^{+1}_{-0.5}$       & 1.03  & 6 \\
Vy 2-2    &$10400^{+600}_{-600}$ & $7000^{+1000}_{-1000}$    & $12^{+4}_{-2}$   & $5^{+1}_{-0.5}$       & 11.80  & 4\\
\enddata
\tablerefs{(1) \citet{LiuXW2001}; (2) \citet{Ruiz2003}; (3) \citet{Tsamis2003}; (4) \citet{Wesson2005}; (5) \citet{ZhangY2005}; (6) \citet{Wang2007}; (7) \citet{Fang2011}; (8)
\citet{Bohigas2015}; (9) \citet{McNabb2016}; (10) \citet{Madonna2017}; (11) \citet{Garcia2018}
}
%\tablecomments{Some date are not in this table because they are out of diagnostic diagram too much. }
%\label{Tresults}
\end{deluxetable}

%\begin{deluxetable}{ccc}
%\tablenum{2}
%\tablecaption{The fitting parameters of Equ.~(\ref{eqfit}) \label{fitting}}
%\tablewidth{10pt}
%\tablehead{
%\colhead{$T_{\rm MB}$([\ion{O}{3}]) (K)} & \colhead{$\alpha$}  & \colhead{$\beta$}}
%\startdata
%10000 & 3.7591  & 0.4854 \\
%12000 & 3.3631  & 0.4197 \\
%14000 & 3.0145  & 0.3461 \\
%\enddata
%\end{deluxetable}

\end{document}